# 50-W average power Ho:YAG SESAM-modelocked thin-disk oscillator at 2.1 µm

**SERGEI TOMILOV,**[1,*] **YICHENG WANG,**[1] **MARTIN HOFFMANN**[1]**, JONAS HEIDRICH**[2]**, MATTHIAS GOLLING**[2]**, URSULA KELLER**[2] **AND CLARA J. SARACENO**[1]

[1]*Photonics and Ultrafast Laser Science, Ruhr Universität Bochum, Universitätsstraße 150, 44801 Bochum, Germany*
[2]*Department of Physics, Institute for Quantum Electronics, ETH Zürich, Auguste-Piccard-Hof 1, 8093 Zürich, Switzerland*
*\*Sergei.Tomilov@ruhr-uni-bochum.de*

**Abstract:** Ultrafast laser systems operating with high-average power in the wavelength range from 1.9 µm to 3 µm are of interest for a wide range of applications for example in spectroscopy, material processing and as drivers for secondary sources in the XUV spectral region. In this area, laser systems based on holmium-doped gain materials directly emitting at 2.1 µm have made significant progress over the past years, however so far only very few results were demonstrated in power-scalable high-power laser geometries. In particular, the thin-disk geometry is promising for directly modelocked oscillators with high average power levels that are comparable to amplifier systems at MHz repetition rate. In this paper, we demonstrate Semiconductor Saturable Absorber Mirror (SESAM) modelocked Ho:YAG thin-disk lasers (TDLs) emitting at 2.1-µm wavelength with record-holding performance levels. In our highest average power configuration, we reach 50 W of average power, with 1.13-ps pulses, 2.11 µJ of pulse energy and ~1.9 MW of peak power. To the best of our knowledge, this represents the highest average power, as well as the highest output pulse energy so far demonstrated from a modelocked laser in the 2-µm wavelength region. This record performance level was enabled by the recent development of high-power GaSb-based SESAMs with low loss, adapted for high intracavity power and pulse energy. We also explore the limitations in terms of reaching shorter pulse durations at high power with this gain material in the disk geometry and using SESAM modelocking, and present first steps in this direction, with the demonstration of 30 W of output power, with 692-fs pulses in another laser configuration. In the near future, with the development of a next generation of SESAM samples for this wavelength region, we believe higher pulse energy approaching the 10-µJ regime, and sub-500-fs pulses should be straightforward to reach using SESAM modelocking.

## 1. Introduction

Average power scaling of ultrafast laser systems operating in the short-wave infrared region (SWIR) from 1.4 µm to 3 µm continues to attract significant interest for a variety of applications. In particular, powerful laser systems operating in the 1.9 µm to 3 µm wavelength region have seen tremendous progress in the last decade, mostly driven by their potential for scientific applications as drivers for XUV sources with high photon energies [1], for the generation of powerful broadband sources for spectroscopy in the mid-infrared region (MIR) (5 µm to 20 µm) [2,3], and for efficient conversion into the terahertz spectral region [4,5], as well as many other areas.

So far, accessing this wavelength region with high power and pulse energy has often been achieved with complex parametric amplifiers pumped by ultrafast laser systems in the more common 1-µm wavelength region [6,7]. A more elegant approach that allows for significantly simpler systems is the use of gain materials directly emitting in this wavelength region, and in laser geometries that allow for power scaling. Fig. 1 shows an overview of high-power ultrafast systems in the range from 1.9 µm to 3 µm, including thin-disk, bulk and fiber systems, illustrating the significant advances in terms of ultrafast power scaling that have been achieved in the last years. Among these systems, some remarkable recent achievements include the highest average power of 1060 W from a Tm-based fiber chirped-pulse amplifier (CPA) system at 1.9 µm [8], as well as 123 W from a coherently-combined CPA system with 0.2-mJ pulse energy at 1.94 µm [9]. For Ho-based systems emitting at 2.1 µm, 142-W continuous-wave (cw) lasing has been achieved from a bulk Ho:YAG system [10], and 55 W with 55-mJ and 4.3-ps pulses from a Ho:YLF regenerative CPA [11]. In addition to amplifiers that provide high average power and high energy, significant efforts have been dedicated to developing bulk solid-state oscillators with different hosts. In this area, the highest power levels achieved so far were demonstrated using Cr:ZnS/ZnSe in multi-W-level, fs-oscillators in the 100-MHz repetition rate region [12].

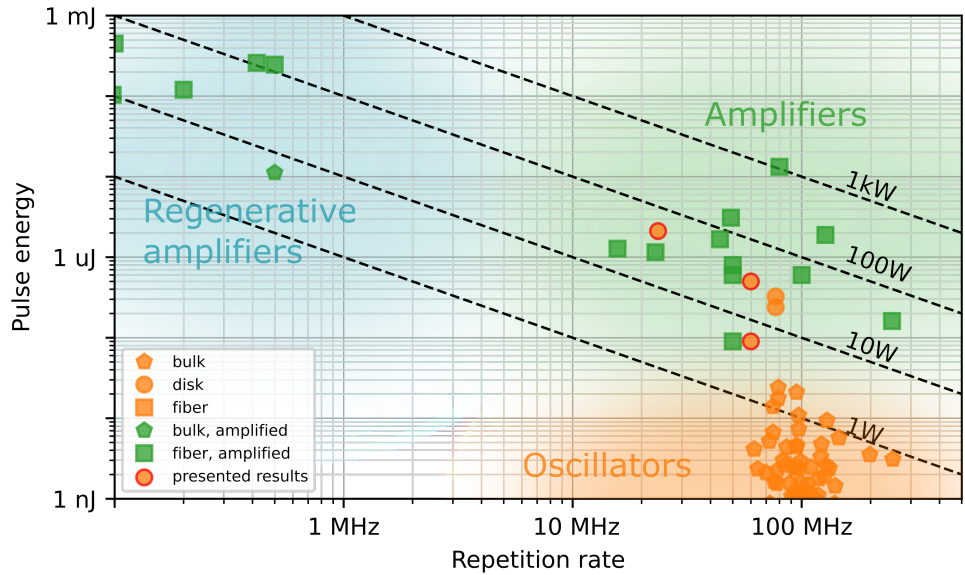

Fig. 1. Overview of 1.9 µm to 3 µm ultrafast laser systems for different active media and geometries in the pulse repetition frequency range from 100 kHz to 500 MHz.

From Fig. 1, one can easily recognize that thin-disk lasers remain widely unexplored in this spectral region, yet offer great potential for energy scaling in amplifier systems, and for directly modelocked, high-power oscillators, as shown by recent achievements in the 1-µm region [13–15]. So far, only two results have conclusively shown this unexplored potential: the demonstration of a Ho:YAG Kerr-lens modelocked single-oscillator TDL with a maximum output power of 25 W and pulse duration of 220 fs [16], and the more recent demonstration of single-mode operation with >100 W of cw power [17].

In this paper, we demonstrate SESAM soliton-modelocked Ho:YAG TDLs emitting at 2.1 µm with record high average power and pulse energy levels. In a double gain-pass configuration, we reach an average power of 50 W with a pulse duration of 1.13 ps at a repetition rate of 23.6 MHz, corresponding to a pulse energy of 2.11 µJ and a peak power of

~1.9 MW. To the best of our knowledge, this represents the highest average power so far demonstrated from a modelocked laser in the 2-µm wavelength region, as well as the highest output pulse energy, delivering output levels that are comparable to amplifier systems at this repetition rate. In a single gain-pass laser configuration, we were also able to demonstrate 30-W output power with 692-fs pulses, which corresponds to 0.48-µJ pulse energy and ~0.7-MW peak power. This record performance level was enabled by the recent development of high-power GaSb-based SESAMs with low loss, adapted for high intracavity power and pulse energy [18]. We believe such high-repetition rate and high-power oscillators will find widespread interest in a variety of applications, for example for powerful oscillator-driven mid-IR generation after subsequent pulse compression, or for material processing applications, which are already within reach using the ultrafast laser system presented here using fast scanning methods.

The obtained results are organized in two parts corresponding to the single gain-pass and double gain-pass configurations that allowed for the above-mentioned different regimes of operation: one with shorter pulses and slightly lower average power, one for record-high power with ps-pulses.

## 2. Single gain-pass results

### 2.1 Fundamental-mode resonator and cw performance

The thin-disk gain medium used throughout the experiment is a 190-µm thick Ho:YAG disk, with a $Ho^{3+}$-doping concentration of 2 at.%, which has a radius of curvature (RoC) of 2 m at room temperature. This disk is pumped by a commercial Tm-fiber laser from IPG Photonics emitting up to 210 W at 1908 nm. In order to efficiently absorb the pump radiation, we use a commercial thin-disk pumping unit (Trumpf GmbH), providing 72 passes of the pump radiation through the active medium, ensuring a high pump absorption of ~97%. The output of the pump collimator was coupled to the pumping unit system through a home-built free-space imaging system.

In Fig. 2, we present the fundamental-mode resonator used throughout the experiments with one single gain reflection. The resonator length in this case was equal to 2.37 m, resulting in a repetition rate for the modelocking experiments of ~63.2 MHz. Using this setup, we first tested the achievable power levels in cw operation with different output coupling mirrors, replacing the SESAM by a high reflector (HR) as an end mirror. It is worth highlighting that the cw tests presented here were performed with all the components needed for achieving modelocking (Brewster plate, dispersive mirrors) in the cavity, in order to have an intracavity loss reference for the modelocking experiments. Without all these components, the achievable cw levels are significantly higher as presented in detail in [17], illustrating one of the difficulties of achieving high modelocking power in this setup.

In order to achieve stable soliton modelocking [19], self-phase modulation (SPM) and negative group delay dispersion (GDD) need to balance each other at every round-trip in the resonator. For precise SPM control we therefore introduced a 3-mm thick plate of sapphire on a translation stage and at Brewster's angle between the mirrors of a telescope, formed by two curved concave mirrors with RoC of 200 mm. By moving this plate closer or further from the telescope waist, one can smoothly change the amount of introduced SPM, due to the mode size change in the plate. In the cw tests shown here, all folding mirrors in the resonator are flat negative-dispersion mirrors, introducing -1000 $fs^2$ of group delay dispersion (GDD) each, and thus providing a total amount of -16000 $fs^2$ of GDD per round-trip. The 3-m RoC convex mirror allows to collimate the laser mode after reflection off the curved disk and therefore, a reasonably large beam size on the critical components is maintained.

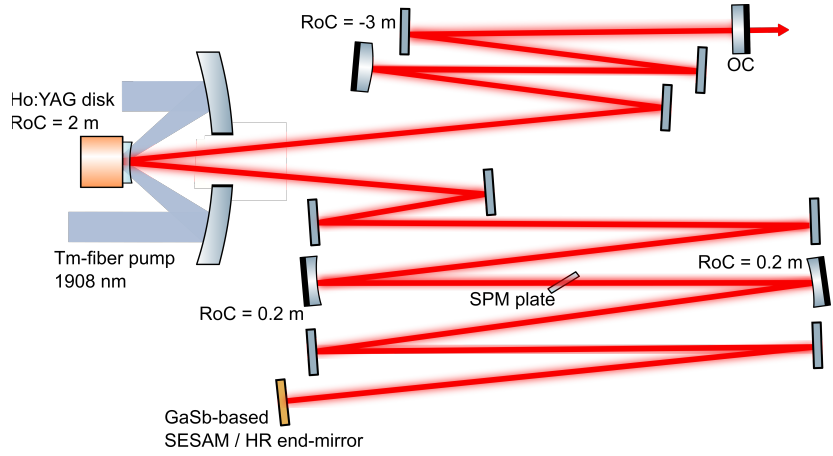

Fig. 2. Single gain-pass modelocked cavity design.

Fig. 3a shows the obtained cw performance (output power and optical-to-optical efficiency) of the laser with different output couplers (OCs) with a 3-mm sapphire plate for SPM. A maximum slope efficiency of 33.9% was achieved with a 3% OC, which is significantly lower than 54.6%, achieved from the simple v-shaped cw cavity presented in [17], indicating a significantly increased level of intracavity losses due to the overall complexity of the cavity. At the intracavity power of ~1.2 kW, measured with a 2% OC, the cw laser was operating close to the fundamental-mode with an $M^2$ of 1.18 and 1.19 along $x$ and $y$-axes (see Fig. 3b), measured by a scanning-slit beam profiler, which is very promising for achieving stable modelocking.

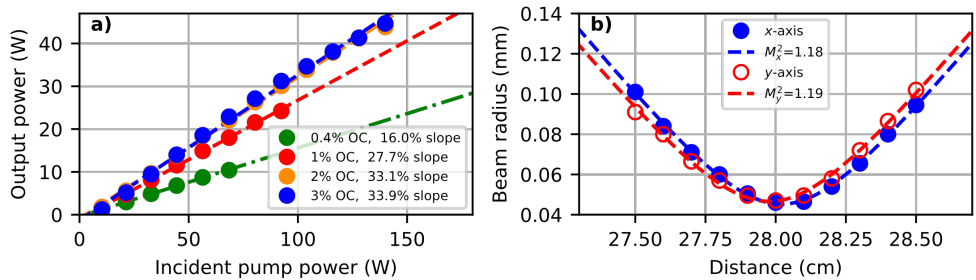

Fig. 3a) cw single gain-pass cavity performance with different OCs and b) beam quality measurement results at ~1.2 kW of intracavity power with a 2% OC.

## 2.2 High-power SESAMs

In order to achieve stable soliton modelocking, a SESAM was used as one of the end-mirrors, as shown in Fig. 2. The GaSb-based SESAM for this laser system was designed and grown using molecular beam epitaxy (MBE) at ETH Zurich [18]. It consists of an absorption section with 9 quantum wells (QWs) on top of a 20-pair distributed Bragg reflector (DBR). These QWs are placed in three antinodes of the standing wave pattern of the incident electric field of the laser. The nonlinear reflectivity parameters of the as-grown SESAM are a saturation fluence $F_{sat}$ of 2.3 µJ/cm², a modulation depth $\Delta R$ of 4.4%, non-saturable losses $R_{ns}$ of 0.17%, and roll-over parameter $F_2$ of 21.3 µJ/cm², measured at 2090 nm. Since this SESAM has a low saturation fluence and we aim here for high intracavity pulse energy, a dielectric top-coating (DTC) is added on top of the absorber section to reduce the field enhancement at the quantum-well positions, resulting in an increased saturation fluence of 6.8 µJ/cm² and a reduced modulation depth of 1%. The parameters of the coated and as-grown samples are collected in Tab. 1:

Tab. 1. Optical properties of coated and uncoated SESAM samples

| Sample | $F_{sat}$ (µJ/cm²) | $\Delta R$ (%) | $R_{ns}$ (%) | $F_2$ (µJ/cm²) |
|---|---|---|---|---|
| as-grown | 2.3 | 4.4 | 0.17 | 21.3 |
| coated | 6.8 | 1 | 0.18 | 213.4 |

The top-coated sample was used for all the high-power modelocking experiments presented in this paper. The non-saturable loss level of the coated sample was measured to be 0.18%, which is remarkably low, and the DTC also allows for improving the damage properties of the sample [20]. The nonlinear reflectivity of the coated sample is shown in Fig. 4. This measurement was performed using a high-precision reflectivity setup [21] adapted for the desired wavelength range [18] with 120-fs pulses at a center wavelength of 2090 nm. It is worth mentioning that the saturation fluence of the sample still remains significantly lower in comparison with SESAMs available for 1-µm TDLs [20], and thus requires very large spot sizes (>1.4-mm radius) for stable operation at high pulse energy. This poses special challenges in the resonator design, the tolerable flatness and lensing of the sample, and thus shows a clear path for further optimizing of such 2-µm SESAMs in the near future [22]. Furthermore, we note that the rollover occurring at high fluences is here enhanced by the short pulse duration of the laser used for SESAM characterization. In our laser, the achieved pulse durations are longer, therefore the rollover fluence becomes larger allowing to operate stably at stronger saturation, as we mention during the description of the laser results.

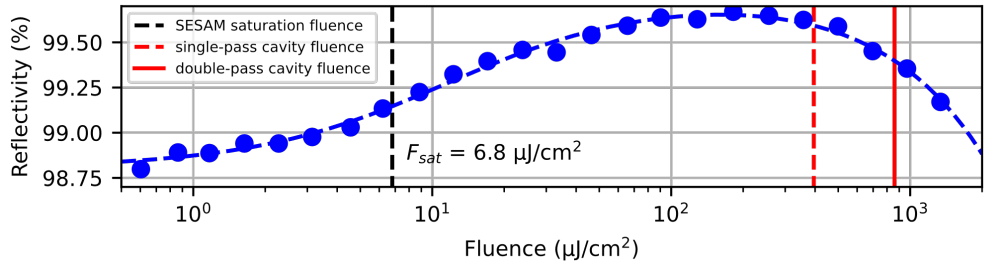

Fig. 4. Nonlinear reflectivity measurement of the coated SESAM together with a fit according to [23] with estimated fluences for achieved laser regimes. Red lines indicate incident fluence estimations for the achieved high-power modelocking regimes.

### 2.3 High-power single gain-pass modelocking performance

Stable soliton modelocking is achieved within a wide range of output powers with 1%, 2% and 3% OC transmission rates with a single gain-pass on the disk. The laser operates stably in the air environment without any cover and temperature stabilization. Here, we present two operation regimes with best performance.

The best power performance was obtained with 2% OC, -16000 fs² of GDD per roundtrip and a 3-mm thick sapphire plate, which provides a phase-shift value γ of $3.4 \cdot 10^{-9}$ W$^{-1}$. In this configuration, we obtained a maximum output power of 30.6 W at a pump power of 95 W, corresponding to a pulse energy of 0.48 µJ at ~63.2 MHz.

Note that there is non-negligible loss or leakage from all cavity mirrors and the Brewster plate, which add up to 6.4 W at the maximum output power, corresponding to 20% of the measured output power from the 2% OC. These sources of loss were, however, calibrated in the cw experiments of section 2.1 and explain the degraded cw performance compared to our record cw results [17]. The fact that the optimal modelocked power was obtained at a lower output coupling rate compared with the cw results is most likely due to increased intracavity losses induced by the small non-saturable loss of the SESAM.

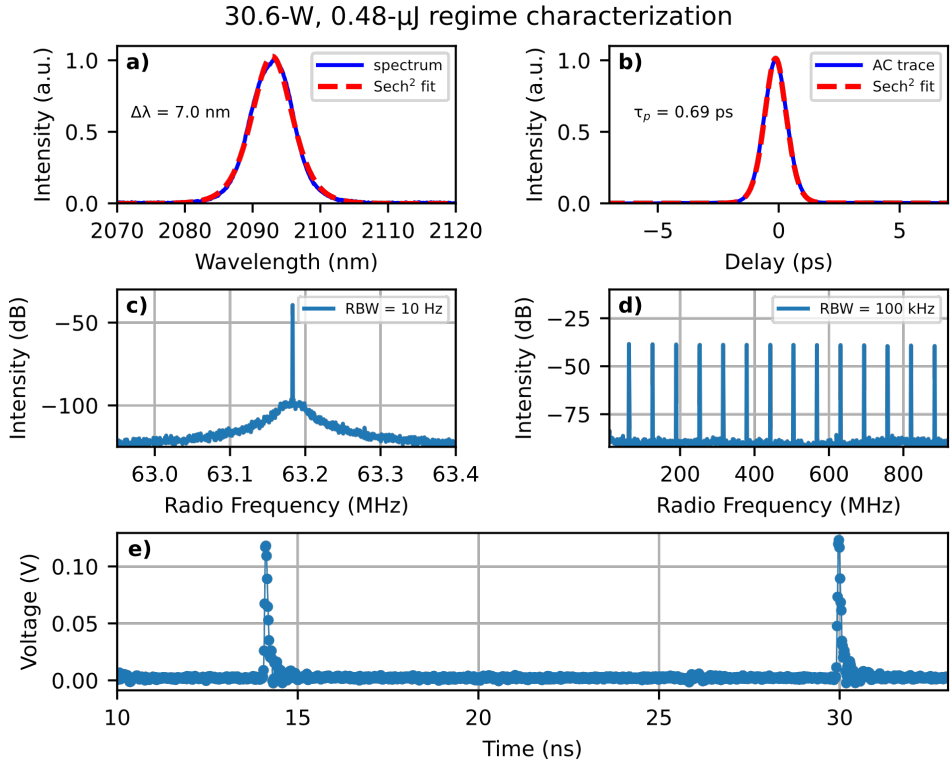

Fig. 5. Characterization of high-power single-pass regime with 30.6 W and 690 fs a) Modelocked laser spectrum b) autocorrelation trace. Radio-frequency spectra of c) first beatnote, d) harmonics in a 900-MHz span, RBW stands for resolution bandwidth. e) Sampling oscilloscope restoration of one repetition period.

The full characterization of the modelocked pulses is shown in Fig. 5. The pulses emitted in this configuration had a spectral Full Width at Half Maximum (FWHM) of 7 nm (Fig. 5a) and 692-fs FWHM pulse duration (Fig. 5b), corresponding to a Time-Bandwidth Product (TBP) of 0.332, which is close to the soliton Fourier limit of 0.315. Fig. 5c shows the measurement of the fundamental repetition frequency using a fast ET-5000 photodetector with a 10-GHz bandwidth and rise/fall times of 28 ps, and Fig. 5d shows the harmonics thereof in a 900-MHz span. The first beatnote is more than 60 dB above the noise level and its harmonics show almost no decay and/or modulations within the long span range, which allows to conclude the absence of Q-switching and short-term instabilities in the signal. To further prove that the laser emits single pulses at the fundamental pulse repetition frequency we used a sampling oscilloscope, which can capture input signal on receiving of a trigger event and reconstruct waveforms with a maximum bandwidth of 25 GHz. Using this device and triggering it with the laser repetition frequency, we can sample the pulse-train with an effective bandwidth limited by the photodiode. Fig. 5e depicts one sampled repetition period of the studied regime and confirms perfect fundamentally modelocked operation of the laser. The difference in peak height is only a consequence of unideal overlap of sampling events and the actual maximum of the signal. This measurement was complemented by checking for eventual cross-correlations in the autocorrelation by scanning the delay over a range of 60 ps. The combination of these two measurements allow us to confirm fundamentally modelocked operation of the laser.

Concerning the obtained output power, any further increase of the pump power leads to an increase of instabilities and corresponding high risk of SESAM damage as well as increased thermal load on other intracavity components due to the extremely high intracavity power of more than 1.5 kW. For the regime described above, we estimate the SESAM fluence to be 395 µJ/cm$^2$, which is already deep into measured roll-over region of the saturation curve according to Fig. 4. However, while the SESAM saturation curve was measured with 120-fs pulses, the obtained Ho:YAG TDL pulse duration is significantly longer. Assuming, that the curve roll-over is dominated by Two-Photon Absorption (TPA), and therefore strongly dependent on the pulse peak intensity, we expect the actual roll-over to happen at significantly higher fluence levels. Therefore, we assume that one of our main limitations at this point is the extreme intracavity power and related thermal effects. As it was also mentioned above, further power-scaling may be possible by increasing the OC transmission while keeping the intracavity power at the same level, however in our case, the laser was not able to reach similar output powers with a 3% OC due to the high intracavity losses.

### 2.4 Short-pulse operation

In a second experiment, we explored the potential limits in the achievable pulse duration. In order to do this, we minimized the intracavity losses by switching to an OC with a transmission of 1% and scanned the amount of intracavity SPM to achieve maximum spectral width at a fixed GDD, which was reduced to -14000 fs$^2$. Detailed results are shown in Fig. 6. The shortest pulse duration achieved was 365 fs at an output power of 5 W with a corresponding phase shift of $1.5 \cdot 10^{-9}$ W$^{-1}$.

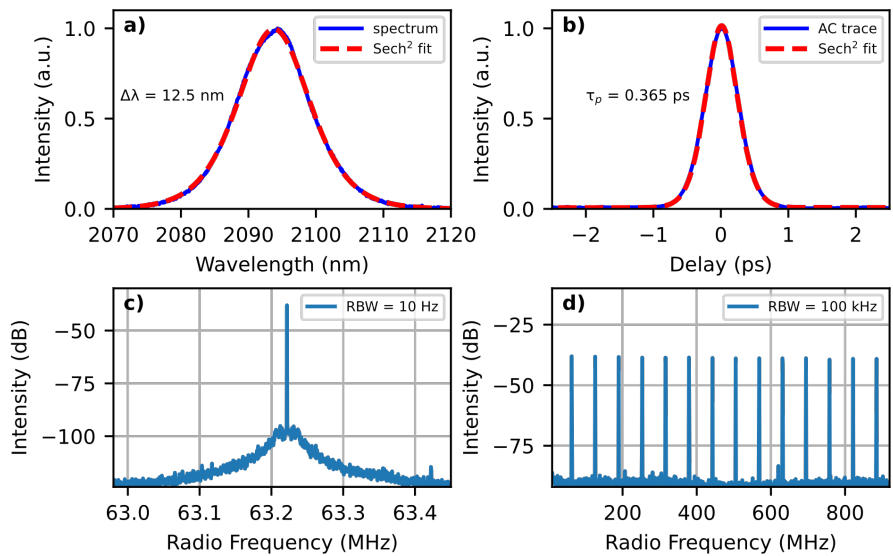

Fig. 6. Short pulse regime a) Modelocked laser spectrum and b) autocorrelation trace. Radio-frequency spectra of c) first beatnote, d) harmonics in a 900 MHz span, RBW stands for resolution bandwidth.

It is worth noting that by keeping the higher OC coefficient of 2%, which should allow to extract more power, these short pulses were not possible to obtain. Our observations generally show a strong tradeoff between OC transmission and shortest achievable pulse duration. This is partly due to the strongly structured gain profile of the Ho:YAG gain [17,24] in combination with the moderate modulation depth available from the used SESAM, and the resulting low tolerable phase shift values. As it can be seen in Fig. 7, the gain profile of Ho:YAG at higher inversion ratios becomes centered around the spike at 2090 nm, and this peak dominates among adjacent spectral components of the structured gain profile. We have already confirmed this spectral behavior for cw laser operation with different OCs in [17] and in modelocked operation, this effect can limit spectral broadening of the pulse if intracavity SPM supported by the SESAM modulation depth is not strong enough to replenish the spectral components that do not experience sufficient amplification per round-trip.

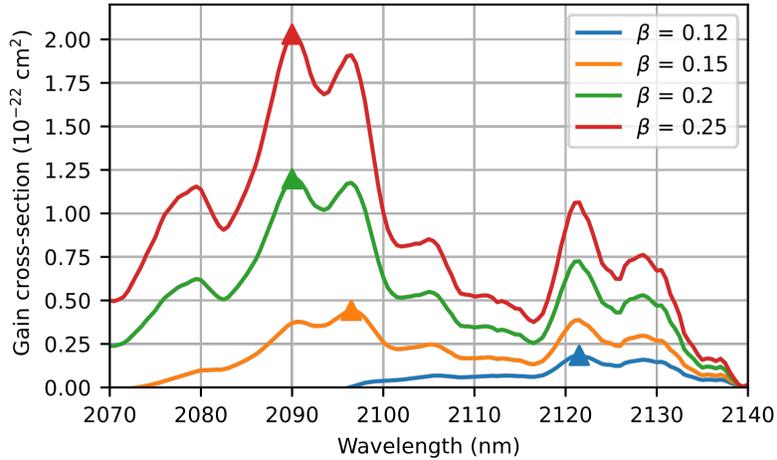

Fig. 7. Gain cross-section over wavelength for different values on population inversion ratio $\beta$ [24]. Triangles indicate maximum gain value for each curve.

For this reason, we were not able to scale the power of this regime by further increasing the OC transmission ratio. We believe that power-scaling of SESAM-modelocked Ho:YAG TDLs while preserving short pulse durations requires a more careful treatment of all intracavity losses, a higher gain (thicker disks, higher doping concentration) and higher modulation depth SESAMs that are compatible with kW intracavity power levels.

## 3. Double gain-pass laser cavity with record-high output power

### 3.1 Resonator design and cw performance

In order to overcome the limitations in the available gain and increase the pulse energy of the modelocked TDL, we implemented a second gain-pass through the disk. The corresponding resonator design is shown in Fig. 8. This double gain-pass design was realized using a 4-*f* telescope extension with the curved disk acting as both telescope mirrors. This cavity has a significantly increased length in comparison with the previous design and therefore its repetition rate is significantly lower – ~24 MHz instead of ~63 MHz, which allows to reach higher pulse energies at the same intra and extra-cavity power levels. In order to accommodate for the higher pulse energy, we increased the RoC of the curved mirrors of the SPM-telescope from 200 mm to 300 mm, thus increasing the beam waist size between them and consequently also reducing the nonlinearities introduced by air. Furthermore, we increased the amount of round-trip GDD from -16000 fs$^2$ to -21000 fs$^2$. We additionally replaced the sapphire plate by an undoped YAG plate, which showed significantly less intracavity losses. All further experiments presented here are carried out with a 1-mm YAG plate.

Fig. 9a shows the cw laser efficiency obtained with different OC coefficients. The optimum OC coefficient of 5% confirms the higher available gain due to the second gain-pass. The $M^2$ parameter measured at ~1.2 kW of intracavity power with 2% OC remains excellent with an $M^2$ of 1.12 and 1.06 along *x* and *y*-axes, respectively.

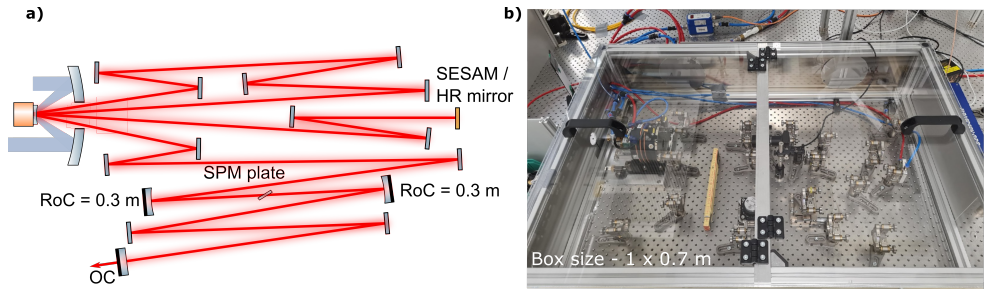

Fig. 8a) Double gain-pass modelocked cavity design, b) actual laser cavity in the non-hermetic box environment.

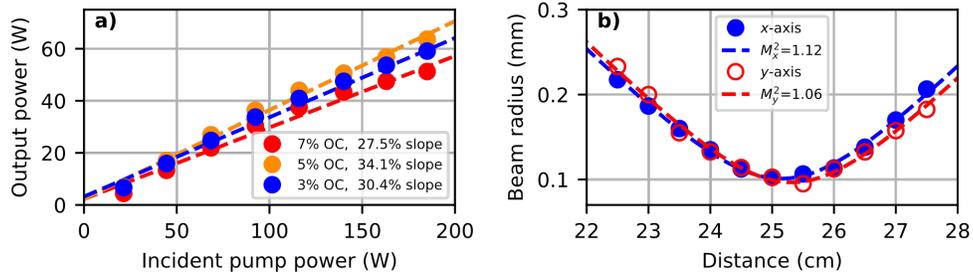

Fig. 9a) cw cavity performance with different OC values and b) beam quality measurement results for double gain-pass cavity at ~1.2 kW of intracavity power with 2% OC.

### 3.2 Modelocked performance

Modelocking of this laser was performed by introducing the same SESAM as we used for the single gain-pass cavity. Stable soliton modelocking was achieved within a wide range of pulse-durations and output powers with OC transmission coefficients in the range from 1% to 4%. Here, we present the regime with the maximum output power of 50 W, corresponding to a pulse energy of 2.11 µJ, achieved at the full pump power of 209 W with a 3% OC.

Fig. 10 shows the full characterization of the modelocked pulses at 50 W of output power. The pulses had a FWHM duration of 1.13 ps with 4.5-nm FWHM spectral bandwidth, which corresponds to a TBP of 0.348. The radio-frequency spectrum also confirms stable modelocking. Single-pulse operation was also confirmed using the same technique as in the single gain-pass case.

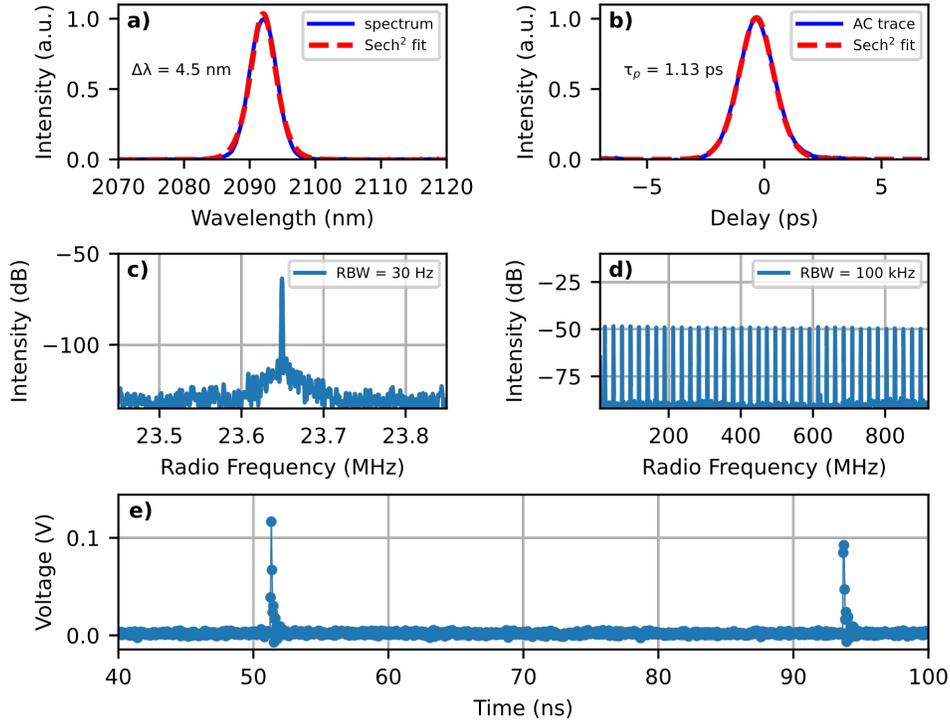

Fig. 10. 50-W regime a) Modelocked laser spectrum b) autocorrelation trace. Radio-frequency spectra of c) first beatnote, d) harmonics in a 900 MHz span. e) Sampling oscilloscope restoration of one repetition period.

The stability of the demonstrated modelocking was affected by air fluctuations due to the increased cavity length of ~6.2 m and because of the high intracavity power. Nevertheless, the laser did not require any re-alignment during characterization.

It is worth to mention, that this regime was achieved at the intracavity power of ~1.6 kW, which is almost the same as in the single-pass cavity 30.6-W modelocking case. The higher gain provided by the second pass allowed us to couple this intracavity power out more efficiently using a 3% OC. Further average power-scaling of the achieved regime is theoretically possible with even higher OC transmission coefficients, if more pump power is available to compensate the growing losses. Another option is to increase the number of gain-passes further to operate at lower intracavity power levels [25].

With 50-W and using a 3% OC, we estimate the fluence on the SESAM to amount to 860 µJ/cm$^2$ which is again reaching the measured short-pulse roll-over region of its saturation curve. Oversaturation of the SESAM is another critical point to consider for further energy scaling, which would be facilitated by a new generation of SESAMs with higher saturation fluence. With respect to the pulse duration, shorter pulses could be reached with higher modulation depths – however this is more critical to combine with the required low loss levels.

We also confirmed the trade-off between pulse duration and intracavity losses for the double gain-pass cavity. We achieved an output power of 7 W in 600-fs and 0.3-µJ pulses using a 1% OC, and 24 W of power in 900-fs and 1-µJ pulses with 2% OC, which can be explained by the same reasons as discussed in section 2.4. External pulse compression techniques would be an excellent alternative to intracavity pulse duration optimization at these comparatively high pulse energy levels.

## 4. Outlook and Conclusion

In summary, we have demonstrated several soliton mode-locked high-power Ho:YAG TDLs. Two oscillator designs were presented – 63.2-MHz single gain-pass and 23.6-MHz double gain-pass cavities. With the single gain-pass cavity, 30.6 W of output power, corresponding to 0.48-µJ pulse energy were achieved at a pulse duration of 692 fs. Additionally, we demonstrated a 5-W regime with a short pulse duration of 365 fs. A longer double gain-pass cavity allowed us to achieve 50 W of output power in 2.11-µJ pulses with a duration of 1.13 ps at the maximum available pump power of 209 W. To the best of our knowledge, this laser system represents the most powerful 2-µm ultrafast oscillator demonstrated so far and further confirms the great potential of $Ho^{3+}$-based TDLs for high-power and high-energy 2-µm laser systems. In the future, we will focus our attention on reducing the pulse duration and increasing the available pulse energy towards the 10-µJ level at this and higher output average power levels. For this, a full exploration of the interplay between the structured gain profile, the SESAM parameters and the obtained power/pulse duration levels will be performed with experiments and simulations. Meanwhile, extra-cavity pulse compression can offer a straightforward path to using this laser system in scientific applications, for example for efficient THz/MIR generation systems for spectroscopy.

**Funding.** Funded by the Deutsche Forschungsgemeinschaft (DFG, German Research Foundation) under Germanys Excellence Strategy – EXC 2033 – Projektnummer 390677874 - RESOLV.

These results are part of a project that has received funding from the European Research Council (ERC) under the European Union's Horizon 2020 research and innovation program (grant agreements No. 805202 - Project Teraqua (Prof. Saraceno) and No. 787097 – Project ONE-MIX (Prof. Keller)).

**Disclosures.** The authors declare no conflicts of interest.

**Data availability.** No data were generated or analyzed in the presented research.